\documentclass[11pt,preprint]{aastex}

\slugcomment{\parbox[h]{5cm}{NAOJ-Th-Ap 2001, No.12}}

\begin{document}
\title{Heating of Intracluster Gas by Jet Activities of AGN: Is the  "Preheating
" Scenario Realistic? }

\author{Masako Yamada\altaffilmark{1} and Yutaka Fujita\altaffilmark{2}}
\affil{National Astronomical Observatory Japan, Osawa 2-21-1, Mitaka,
Tokyo, 181-8588, Japan} 
%\email{masako@th.nao.ac.jp}

\altaffiltext{1}{email:masako@th.nao.ac.jp}
\altaffiltext{2}{Department of Astronomy, University of Virginia, P.O. Box 3818, Charlottesville,
 VA 22903-0818; email:yf4n@astsun.astro.virginia.edu,
}

%----\input{abst}
\begin{abstract}
We investigate the non-gravitational heating of hot gas in clusters of
galaxies (intracluster medium; ICM) on the assumption that the gas is
heated well before cluster formation ('preheating'). We examine the jet
activities of radio galaxies as the sources of excess energy in ICM, and
the deformation of the cosmic microwave background (the
Sunyaev-Zel'dovich effect) by hot electrons produced at the jet terminal
shocks. We show that the observed excess entropy of ICM and {\sl
COBE/FIRAS} upper limit for the Compton $y$-parameter are compatible
with each other only when the heating by the jets occurred at relatively
small redshift ($z\lesssim 3$). Since this result contradicts the
assumption of 'preheating', it suggests that the heating occurred
simultaneously with or after cluster formation.
\end{abstract}


%-----\input{intro}
\section{Introduction}

The departure of the properties of X-ray emitting gas in galaxy clusters
(intracluster medium; ICM) from simple scaling relations gives rise to
arguments about its thermal history
\citep{kaiser86,kaiser91,evrad91,fujita00}.  Observed relations between
X-ray luminosity and temperature show that from a rich cluster scale to
a poor cluster scale, the exponent increases from $ L_X\propto
T_X^{2-3}$ \citep[e.g.][]{david93,xue00} to $L_X\propto T_X^{5}$ \citep
{ponman96,xue00}, which are much steeper than those obtained by
gravitational collapse alone ($L_X\propto T_X^{2}$). Moreover, the
discovery of the excess entropy in poor clusters (``entropy-floor'') by
\citet {ponman99} is presently interpreted as a strong evidence for the
existence of non-gravitational heating in the ICM.

One of the popular scenarios to successfully explain these thermal
properties of ICM is a so-called "preheating" scenario
\citep[e.g.][]{tozzi99}. In this scenario, it is assumed that  the
proto-ICM
is heated {\em well before} the collapse of clusters, which explains the
break in the $L_X-T_X$ relation and the entropy floor as follows. If the
virial temperature of a cluster, $T_{\rm vir}$, is much higher than the
temperature of the external gas, $T_{\rm ex}$, the external gas accreted
by the cluster is heated to $T_{\rm vir}$ by shock waves forming at the
collapse, and the scaling relation $L_{X}\propto T_{X}^2$ should be
satisfied \citep[e.g.][]{cav98}. On the other hand, if $T_{\rm vir}$ is
comparable to or smaller than $T_{\rm ex}$, shock waves do not form at
the collapse, and the external gas adiabatically accretes onto the
cluster. Thus, the gas temperature is determined not only by $T_{\rm
vir}$ but also by $T_{\rm ex}$.  Several authors have shown that
preheating models can reproduce the observational results
\citep[e.g.][]{tozzi99}, however, the heat source itself and the input
epoch have not been identified.

Some authors have investigated the heating by supernovae.  However,
\citet{valageas99} showed that the energy provided by supernovae cannot
raise the entropy of intergalactic medium (IGM) up to the level required
by current observations. Moreover, \citet{kravtsov00} estimated the
energy provided by supernovae from the observed metal abundance of ICM
and found that the heating by supernovae alone requires unrealistically
high efficiency. On the other hand, AGNs may be much more powerful and
are, therefore,
plausible candidates of heating sources, and thus we
focus on AGNs.

As for the input epoch, there have been few concrete arguments except
for the ones inherent in models of individual sources.  For AGNs, there
have been no additional constraints like the metal abundance in the case
of supernova heating \citep{kaiser99}. In this paper, we propose a new
approach to study this subject.
%way to shed a light on this subject.
Preheating sources which are so
powerful as to influence the thermal properties of ICM would also
 deform the
spectra of the cosmic microwave background (CMB) via inverse-Compton
scattering (Sunyaev-Zel'dovich effect; hereafter SZ).  Indeed, it has
been shown that the energy supplied by AGN jets could be a significant
source of SZ effect \citep{yamada99}.  If the preheating scenario is
correct, then the
 ICM should have been heated before the collapse of poor clusters.
This allows us to obtain a lower limit of the redshift of energy input.
%and thus we obtain the lower bound of the energy input redshift.
On the
other hand, if a large amount of hot IGM exists too early, the
cumulative SZ effect would break the constraint by {\sl COBE/FIRAS}
($y\lesssim 1.5\times 10^{-5}$, \citealt{fixen96}). In this paper, we
compare the estimated SZ effect and the excess energy induced by cocoons
formed by AGN jet activities with the observational constraints, and
discuss the validity of the simple preheating scenario.

\section{Models}
%-----\input{cocoon model&y-parameter}
\subsection{Cocoon model and the Compton $y$-parameter}
We assume for simplicity that the heating of proto-ICM by jet activities
occurred well before cluster formation. The
kinetic energy of the jet is
transferred to thermal energy of the proto-ICM via thermalization at the
shock at the hot spot. The thermalized jet matter expands into the
intergalactic medium (or proto-ICM) surrounding the radio galaxy
laterally as well as along the jet axis
\citep{begelman89,nath95,kaiser97,yamada99}. As this hot matter expands
supersonically, a hot region surrounded by a shock surface around the
radio galaxy is expected to form; hereafter we refer to it as a
"cocoon".  We briefly summarize the evolution of the cocoon below
\citep[see for details,][]{yamada99}.

While the jet is active, we can write pressure inside the cocoon as
\begin{equation}
P_c=\left\{ \frac{5}{8}\left[\frac{L_j(\gamma-1)}{\epsilon_vt^2}\right]^2
       \frac{\rho_a^2P_a}{c_a^2}\sin^2\phi \right\}^{1/5},
\end{equation}
by balancing the shock thrust and ram-pressure of the background matter,
and by using Rankine-Hugoniot conditions. In this equation $L_j$
represents the kinetic energy of the jet, $\gamma=5/3$ is the adiabatic
index, $\epsilon_v$ is the "volume factor" which describes the shape of
the cocoon (compared with a sphere), $t$ is the time elapsed since the
ignition of the jet, $\phi$ is the opening angle of the bow shock ahead
of the jet termination spot (see Fig.~1 of \citealp{yamada99}), $c_a$ is
the sound speed, $\rho_a$ is the gas density, $P_a$ is the pressure, and
the suffix $a$ denotes the ambient matter, respectively. In the
derivation of the above equation, we assumed that $L_j$ is
time-independent, and is given by the Eddington luminosity of the
central black hole which activates the jet. We use the gas density of
the background universe as $\rho_a$.
We adopt the black hole mass $M_{\rm BH}=0.002M_{\rm sph}$, where
$M_{\rm sph}$ is the mass of the spheroidal component of the host galaxy
\citep{magorrian98}. We can estimate the total thermal energy deposited
during the jet active phase as $P_cV=\epsilon_{\rm ff}\times (\gamma-
1)L_jt_{\rm life}$, where $V$ is the volume of the cocoon,
$\epsilon_{\rm ff}$ is the thermalization efficiency, and $t_{\rm life}$
is the lifetime of the jet activity.  We adopt the standard value for
$t_{\rm life}$ ($3\times 10^7$ years), and take the thermalization
efficiency for a free parameter.

While the jet lifetime is short, the cocoon can stay hot for about its
cooling time even after the energy supply stops.
This "residual" phase contributes much more
strongly to the SZ effect \citep{yamada99}. We model the
evolution of the cocoon adopting the analogy with the evolution of a
supernova remnant (SNR) in interstellar medium.
Thus we define the effective lifetime of the cocoon as the time from the
death of the jet to the epoch when the expansion time $\tau_{\rm
ex}=R_s/v_ s$ equals the cooling time behind the shock front, $\tau_{\rm
cool}|_{\rm shock}$ ($R_s$ and $v_s$ are the shock radius and the
velocity, respectively).

Finally we write the evolution of the internal energy of a single
cocoon:
\begin{eqnarray}
    P_cV = \cases{ \epsilon_{\rm ff} L_j(\gamma -1)(t-t_{\rm begin}), & $t<
t_{\rm life}$, \cr
                            \epsilon_{\rm ff} L_j(\gamma-1)t_{\rm life}\times
                                  \exp[-(t-t_{\rm life}-t_{\rm begin})/t_r(
z)],
                                   &  $t>t_{\rm life}$, \cr }
\end{eqnarray}
where $t$ is the cosmological time, $t_{\rm begin}$ is the epoch of jet
ignition, and $t_r(z)$ is the effective lifetime at redshift $z$,
respectively.  Numerical simulations have shown that, although
the thermal energy of a SNR rapidly decreases soon after the radiative
phase begins, the cooling time increases as the SNR expands further and
about 10\% of the initial thermal energy is left behind
\citep{chevalier74,thornton98}. Thus we keep the total internal energy
constant after it drops to 10\% of its initial value.

%.... y-estimation

The Compton $y$-parameter is calculated by integrating the product of
total internal energy within a single cocoon and the number of radio
galaxies in the line of sight,
 \begin{equation}
 y  \approx \int\int \frac{P_cV}{m_ec^2}\sigma_Tn_{\rm RG}(M,z_{\rm coll})a
^3r^2dr\frac{1}{R_A^2}dM,  \label{eq:y}
 \end{equation}
where $m_e$ is the electron mass, $\sigma_T$ is the Thomson scattering
cross section, $n_{\rm RG}$ is the comoving number density of radio
galaxies, $z_{\rm coll}$ is the typical collapse epoch of host
galaxy halos, $a$ is the cosmological scale factor, $r$ is the comoving
radial coordinate, $R_A$ is the angular diameter distance, respectively
\citep{yamada99}.
We assume that radio galaxies reside in halos with the mass of
$M>10^{10}\;M_{\sun}$. We also assume that a fraction of normal
galaxies has jet activity, and set the proportional constant to be a
canonical value $f_r=0.01$. We count the number of radio (or normal)
galaxies using the Press-Schechter number density ($n_{\rm PS}$), and do
not use the luminosity function of radio galaxies; this is because we
intend to count up the "residual" cocoons, whose
 effective lifetime
($t_r$) is much longer than the synchrotron-decay time.
%
%for $10^{12}\;M_{\sun}<M<10^{14}\;M_{\sun}$ and $f_r=0.01$ for
%others.  In the Press-Schechter formalism we use the fitting formulae of
%\citet{kitayama96} for the variance of density perturbation $\sigma(M)$,
%\begin{equation}
%    \sigma\propto (1+2.208m^p-0.7668m^{2p}+0.7949m^{3p})^{-2/(9p)},
%\end{equation}
%where p=0.0873 and $m\equiv M(\Gamma h)^2/(10^{12}M_{\sun})$.
%
We define the epoch of jet ignition $t_{\rm begin}$ separately from the
"typical collapse time" $t_{\rm coll}$ of the dark halo of the host
galaxy, at which the variance of density perturbation of scale $M$,
$\left\langle \left(\frac{\delta M}{M}\right)^2 \right\rangle$, is equal
to $1.69^2$ when $\Omega_0=1$.
%
%\begin{equation}
%    \left\langle \left(\frac{\delta M}{M}\right)^2 \right\rangle (M,z)
%    =\frac{1}{2\pi^2}\int^{\infty}_{0}\frac{dk}{k}k^3
%   \langle |\delta_k|^2 \rangle
%    \left[ \frac{3(\sin(kr_M)-(kr_M)\cos(kr_M))}{(kr_M)^3} \right]^2,
%\end{equation}
%
%In this equation, $\delta_k$ is the density
%fluctuation in the wave number space.
We take $t_{\rm gap}\equiv t_{\rm begin}-t_{\rm coll}$ as a free
parameter because of the uncertainty inherent in the jet activation
mechanism.  The value of $t_{\rm gap}$ varies between
0 (jet ignition
coeval with the collapse of the dark halo of its host galaxy)  and
$\lesssim 10^{10}$ years ($\approx H_0^{-1}$).

%-----\input{preheat}
\subsection{Energy Input into Proto-ICM}

It is reasonable to assume that the density of
proto-ICM traces that of galaxies before the density perturbation
corresponding to a protocluster ($\delta$) goes nonlinear.
Thus the local number density of
radio galaxies and the gas density
of proto-ICM in a proto-cluster region are written
as:
\begin{eqnarray}
    n_{\rm gal}(M,z) &=& n_{\rm PS}f_r\left(\frac{a_0}{a}\right)^3(1+\delta),
\label{ecocoon}\\
    n_{\rm gas}(M,z) &=& \frac{\rho_{\rm crit}}{\mu m_p}\Omega_bf_c(1+\delta),
\label{ngas}
\end{eqnarray}
where $a_0$ is the present scale factor, $\rho_{\rm crit}$ is the
critical density, $\mu= 0.59$ is the mean molecular weight for
primordial gas, $m_p$ is the proton mass, and $f_c$ is the fraction of
the gas compared with the baryon density $\Omega_b$, respectively.

According to equation (\ref{ecocoon}), the energy density ejected by
AGNs into a proto-cluster is given as
\begin{equation}
    \epsilon_{\rm tot}=\int_{\rm M_l}f_rn_{\rm PS}(M,z)
            \cdot P_cV(M,z)
            \cdot(1+\delta)dM.
\end{equation}
%
%where $n_c$ is the average number density of ICM particles,
%and
Hence the energy input per nucleon $E_{\rm input}$ at present ($z=0$) is
\begin{equation}
    E_{\rm input} = \frac{\epsilon_{\rm tot}V_c}{n_{\rm gas}V_c}
    = \left(\frac{\rho_{\rm crit}}{\mu m_p}\right)^{-1}
           \frac{1}{\Omega_bf_c}\int_{M_l}n_{\rm PS}(M,0)f_r
               \cdot P_cV(M,0)dM,
\label{eq:einput}
\end{equation}
which measures the additional, non-gravitational heating.  Hereafter we
assume $f_c=1$, which is reasonable when the density contrast is in a
linear regime.
%

%----\input{result}
%%%%%%%%%%%%%%%%%%%%%%%%%%%%%%%%%%%%%%%%%%%%%
\section{Results}

\begin{figure}[h]
  \begin{center}
    \epsscale{0.8}
    \plotone{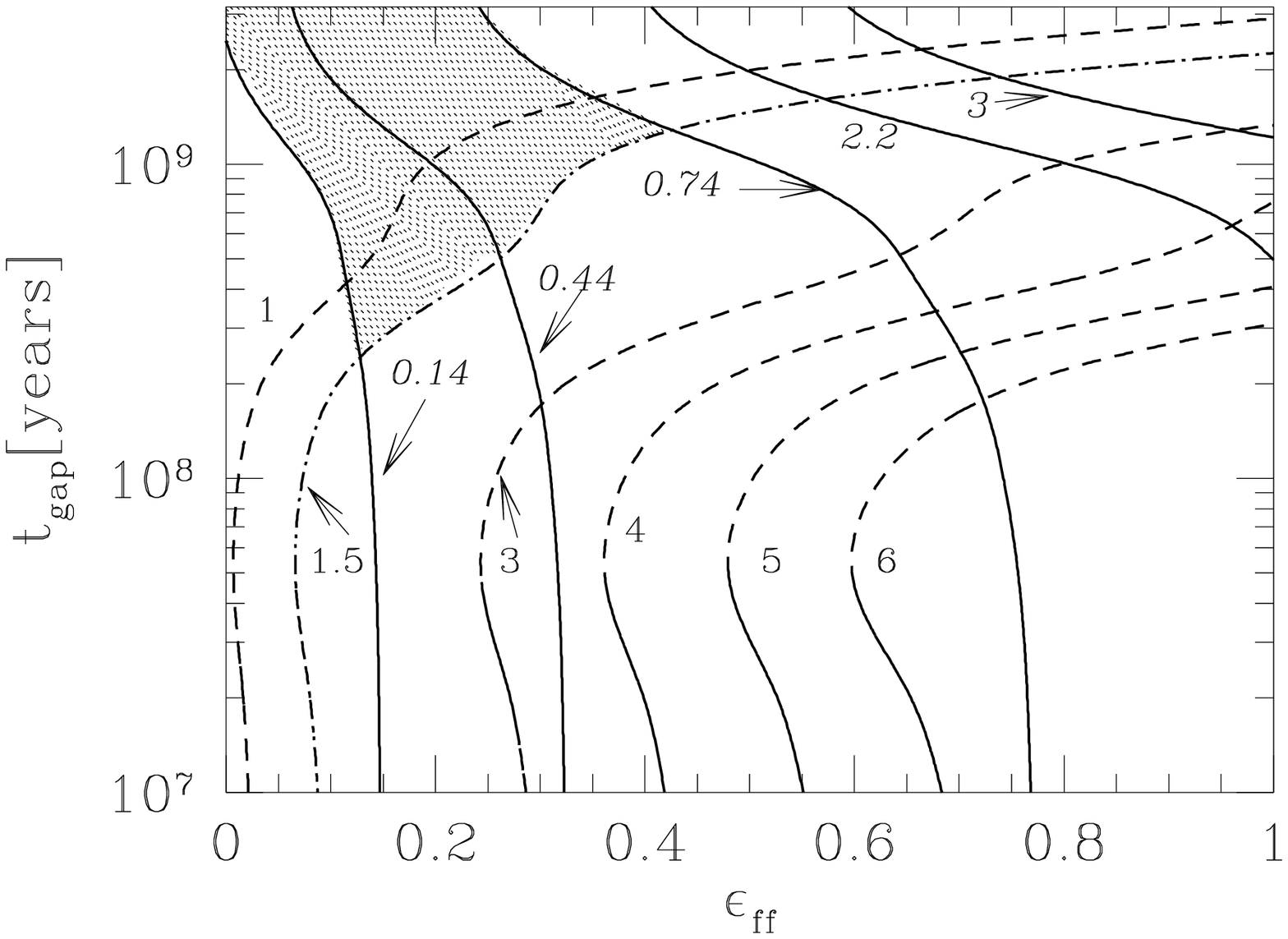}
\end{center}
\figcaption[fig1.eps]{A contour map of the Compton $y$-parameter (dashed
lines) and $E_{\rm input}$ (solid lines). Each line corresponds to
$y/10^{-5}$=1, 1.5, 3, 4, 5, 6 and $E_{\rm input}$=0.14, 0.44, 0.74.
2.2, 3 keV, as indicated in the figure. A shaded region is allowed one
by {\sl COBE} observation ($y\le 1.5\times 10^{-5}$) and by the
non-gravitational heating obtained by \citet{lloyddavies00} (0.44 $\pm
$0.3 keV).  \label{fig1}} 
\end{figure}

We calculate equations (\ref{eq:y}) and (\ref{eq:einput}) with various
combinations of two parameters $\epsilon_{\rm ff}$ and $t_{\rm gap}$. We
adopt the standard cold dark matter cosmology, with $\Omega_0=1, h=0.8$, baryon
density $\Omega_b h^2=0.0125$, and {\sl COBE} normalization
for the density perturbations. In Figure~\ref{fig1} contours of the
Compton $y$-parameter and $E_{\rm input}$ are plotted.
X-ray observations show that excess energy via non-gravitational heating
is 0.44$\pm$ 0.3 keV per a particle \citep*{lloyddavies00}. The
region
where parameter values are consistent with the $y$-parameter
constraint obtained
by
{\sl COBE}, $y\lesssim1.5\times 10^{-5}$ \citep{fixen96}, and the energy
deduced by \citet{lloyddavies00} is indicated as a shaded region.  As is
clearly seen, almost horizontal contours from
the $y$-parameter constraint severely
limit the value of $t_{\rm gap}$ to be $\gtrsim 6.3\times 10^8$ years.

\begin{figure}[h]
  \begin{center}
    \epsscale{0.8}
    \plotone{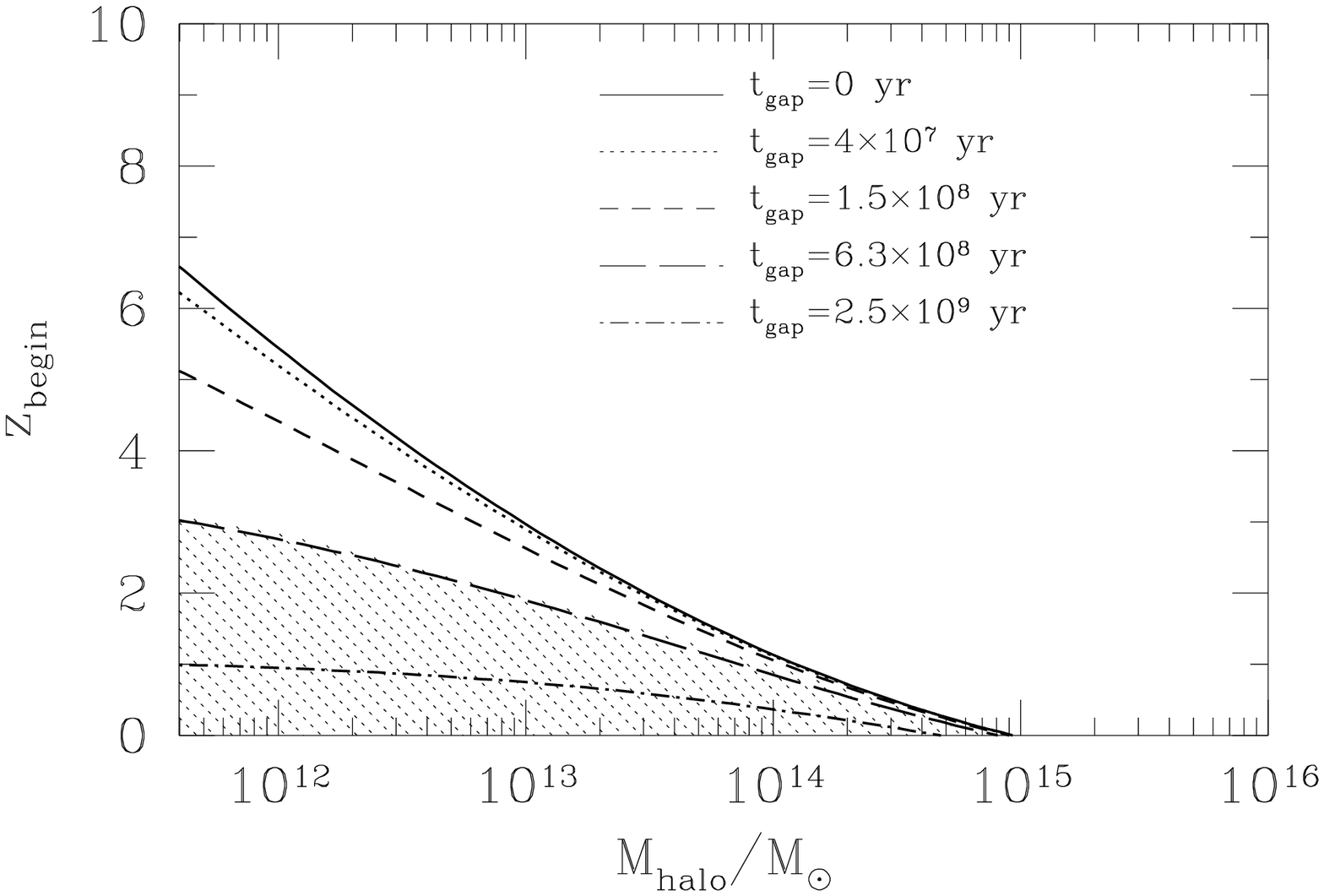}
\end{center}
\figcaption[fig2.eps]{Jet ignition epoch as a function of dark halo mass
of a radio galaxy with various values of $t_{\rm gap}$.  A shaded region
corresponds to the allowed value obtained from the Figure~1 ($t_{\rm
gap}\gtrsim 6.3\times 10^8$ years). Solid line ($t_{\rm gap}=0$)
indicates the typical formation epoch of objects of mass $M_{\rm halo}$.
\label{fig2}}
\end{figure}

The result that the value of $y$ is almost independent of thermalization
efficiency $\epsilon_{\rm ff}$ comes from the weak dependence of the
expansion speed on the internal energy in the "Sedov" phase ($v_s\propto
E_0^{1/5}$, \citealt {shu92}), which results in the weak dependence of
cooling time (effective lifetime of a cocoon) on $\epsilon_{\rm ff}$.
On the other hand, contours of $E_{\rm input}$ limit the thermalization
efficiency to $\epsilon_{\rm ff}\lesssim 0.4$.  The fact that the
contours of $E_{\rm input}$ run vertically for small values of
 $t_{\rm gap}$
reflects the rapid cooling of cocoons at high redshift. When $t_{\rm
gap}$ is small, the cooling time of a cocoon is so short that the cocoon
energy rapidly reduces to 10\% of the energy supplied by the jet; thus
$E_{\rm input}$ is simply given by $L_j t_{\rm life}\epsilon_{\rm
ff}\times 0.1$ and is proportional to $\epsilon_{\rm ff}$.
For late input (large $t_{\rm gap}$) the contours slightly curve to
left; this is due to the small number of cocoons that contribute to the
heating, which means that a large value of $\epsilon_{\rm ff}$ is needed
to gain the same amount of energy.
In order to assign $t_{\rm gap}$ and the jet ignition epoch, we plot the
corresponding redshift $z_{\rm begin}$ as the function of halo mass in
Figure~\ref{fig2}. Figure~\ref{fig1} shows that $t_{\rm gap}\gtrsim
6.3\times 10^8$ years, which means heating of ICM occurred at $ z_{\rm
begin}\lesssim 3$ (Figure~\ref{fig2}). Compared to the halo collapse
epoch ($t_{\rm gap}=0$; solid line), the upper bound is very close to
the formation epoch of poor clusters with the mass of
$M=10^{13}-10^{14}M_{\sun} $.  This suggests that the heating of the ICM
occurred simultaneously with or after the collapse of the poor clusters,
which contradicts the assumption about the heating epoch inherent in the
scenario.  In other words, our results bring up a serious question about
the genuine preheating model.

%

%----\input{conclusion}
\section{Discussion and Conclusions}
We have proposed a new way to elucidate the thermal history of ICM. We
have calculated the Compton $y$-parameter and the energy ejected into
the ICM through AGN jet activities assuming that the non-gravitational
heat input by the jets occurred well before cluster formation. Comparing
them with observations, we have found that the heat input had {\em not}
occurred at $z\gtrsim 3$. Since $z\sim 3$ is the typical formation epoch
of poor clusters, this is not consistent with the assumption of
`preheating' and suggests that the heat input occurred simultaneously
with or after the formation of poor clusters. Considering the wide range
dispersion in the formation epochs of poor clusters \citep[$z\lesssim
3$;][]{lacey93,kitayama96a,balogh99}, the scenario of heating
 coeval with the
cluster formation may be more plausible and would be consistent with the
%concerned in the
dispersions of cluster properties \citep{fujita00}. Note that if the
heating occurred in dense environment like in clusters, the lifetime and
filling-factor of cocoons will decrease \citep{yamada99}, which may reduce
the expected value of the $y$-parameter well below the observational
limit.
\citet{valageas99} also estimated the $y$-parameter for QSO heating, but
found a smaller value than ours ($y\lesssim 10^{-6}$).
%This difference comes from the difference in
%the effective lifetime of QSOs
%or cocoons as the sources of the SZ effect.
%, which results in a larger $y$  than that of \citet{valageas99}.
This may be because in their model, the energy injected into IGM by QSOs
started to cool immediately after the energy injection, which leads to a
smaller $y $ value.  On the other hand, in our model, a cocoon remains
hot for a long time until the cooling sets in (see \S2).

Below we discuss several points which are omitted in our simple model.
First, a part of electrons may be accelerated at the terminal shock to
become
 non-thermal populations. The SZ effect concerning these populations
was calculated by several authors \citep[see e.g.,][]
{birkinshaw99,ensslin00}, and is shown that the amplitude of the signal
is reduced by only a small factor for a fixed total energy of the
gas. Thus, we do not think that the results in our paper change
significantly. Second, although recent works showed that jet matter is
suggested to be electron-positron dominated at least at close to the
core \citep[e.g,][]{wardle98}, some authors proposed that protons
constitute a part of the jet energy \citep[e.g][]{mannheim98}.  If this
is the case, the $y$ value due to the jet matter is accordingly reduced,
and the input energy epoch may not be constrained by SZ effect.
%Energetics
%of jets and radio galaxies, including the above processes have not yet
%fully understood, and still require detailed discussions.

Recently, {\sl Chandra} observations found that there is no indication
of shocks around several radio galaxies in the center of clusters
\citep{ mcnamara00,fabian00}.  These sources reside in such high
pressure regions that the expansion speed of lobes is subsonic \citep[
e.g.][]{fujita01}.  We have considered radio galaxies in proto-ICM which
has not fully collapsed to have such high pressure, and then have shown
that this assumption is not compatible with observational constraints of
$y$ and $E_{\rm input}$.  Thus {\sl Chandra} findings do not alter our
main conclusion.

\acknowledgements We thank useful comments of N. Sugiyama and
M. Nagashima. We greatly appreciate B. Nath for reading this paper critically.

%----\input{bib.tex}


\begin{thebibliography}{}
\bibitem[Balogh et al.(1999)Balogh, Babul, and Patton]{balogh99}
Balogh, M. L., Babul, A., \& Patton, D. R. 1999, \mnras, 307, 263.
\bibitem[Begelman \& Cioffi(1989)]{begelman89}
Begelman, M. C., \& Cioffi, D. F. 1989, \apj, 345, L21.
\bibitem[Birkinshaw(1999)]{birkinshaw99}
Birkinshaw, M. 1999, Phys. Rept. 310, 97.
\bibitem[Cavaliere, Menci \& Tozzi(1998)]{cav98} Cavaliere,
A., Menci, N.\ \& Tozzi, P.\ 1998, \apj, 501, 493
\bibitem[Chevalier(1974)]{chevalier74}
Chevalier, R. A. 1974, \apj, 188, 501.
\bibitem[David et al.(1993)]{david93} David, L.\ P., Slyz, A.,
Jones, C., Forman, W., Vrtilek, S.\ D.\ \& Arnaud, K.\ A.\ 1993, \apj, 412,
479
\bibitem[Ensslin \& Kaiser(2000)]{ensslin00}
Ensslin, T. A., \& Kaiser, C. R. 2000, \aap, 360, 417.
\bibitem[Evrad \& Henry(1991)]{evrad91}
Evrad, A. E., \& Henry, J. P. 1991, \apj, 385, 95.
\bibitem[Fabian et al.(2000)]{fabian00}
Fabian, A. C. et al. 2000, \mnras, 318, L65.
\bibitem[Fixen et al.(1996)]{fixen96}
Fixen, C. J., Cheng, E. S., Gales, J. M., Mather, J. C., Shaffer, R. A., \&
Wright, E. L. 1996, \apj, 473, 576.
\bibitem[Fujita(2001)]{fujita01}
Fujita, Y. 2001, \apjl, Letters, in press. (astro-ph/0102221)
\bibitem[Fujita \& Takahara(2000)]{fujita00}
Fujita, Y., \& Takahara, F. 2000, \apj, 536, 523.
%\bibitem[Hardcastle and Warrall(2000)]{hardcastle00}
%Hardcastle, M. J., \& Worrall, D. M. 2000, astro-ph/0007260.
\bibitem[Kaiser \& Alexander(1997)]{kaiser97}
Kaiser, C. R., \& Alexander, P. 1997, \mnras, 286, 215.
\bibitem[Kaiser \& Alexander(1999)]{kaiser99}
Kaiser, C. R., \& Alexander, P. 1999, \mnras, 305, 707.
\bibitem[Kaiser(1986)]{kaiser86}
Kaiser, N. 1986, \mnras, 219, 785.
\bibitem[Kaiser(1991)]{kaiser91}
Kaiser, N. 1991, \apj, 383, 104.
\bibitem[Kitayama \& Suto(1996)]{kitayama96a}
Kitayama, T., \& Suto, Y., 1996, \mnras, 280, 638.
%\bibitem[Kitayama and Suto(1996b)]{kitayama96b}
%Kitayama, T., \& Suto, Y., 1996, \apj, 469, 480.
%\bibitem[Kitayama and Suto(1997)]{kitayama97}
%Kitayama, T., \& Suto, Y. 1997, \apj, 490, 557.
\bibitem[Kravtsov \& Yepes(2000)]{kravtsov00}
Kravtsov, A. V., \& Yepes, G. 2000, \mnras, 318, 227.
\bibitem[Lacey \& Cole(1993)]{lacey93}
Lacey, C., \& Cole, S. 1993, \mnras, 262, 627.
%\bibitem[Lacey and Cole(1994)]{lacey94}
%Lacey, C., \& Cole, S. 1994, \mnras, 271, 676.
%\bibitem[Leahy and Gizani(1999)]{leahy99}
%Leahy, J. P., \& Gizani, N. A. B. 1999, 'Life Cycles of Radio galaxies', ed.
%J. Biretta et al. (New Astronomy Reviews).
%\bibitem[Liu et al.(1992)Liu, Pooley, and Riley]{liu92}
%Liu, R., Pooley, G., \& Riley, J. M. 1992, \mnras, 257, 545.
\bibitem[Lloyd-Davies et al.(2000)Lloyd-Davies, Ponman \& Cannon]{lloyddavies00}
Lloyd-Davies, E. J., Ponman, T. J., \& Cannon, D. B. 2000, \mnras, 315, 689
%\bibitem[Loewenstein(2000)]{loewenstein00}
%Loewenstein, M. 2000, \apj, 532, 17.
\bibitem[McNamara et al.(2000)]{mcnamara00}
McNamara, B. R. et al. 2000, \apj, 534, L135.
\bibitem[Magorrian et al.(1998)]{magorrian98}
Magorrian, J. 1998, \aj, 115, 2285.
\bibitem[Mannheim(1998)]{mannheim98}
Mannheim, K. 1998, Science, 279, 684.
\bibitem[Menci \& Cavaliere(2000)]{menci00}
Menci, N., \& Cavalier, A. 2000, \mnras, 311, 50.
\bibitem[Nath(1995)]{nath95}
Nath, B. B. 1995, \mnras, 274, 208.
\bibitem[Ponman et al.(1996)]{ponman96}
Ponman, T. J., Bourner, P. D., Ebeling, H., \& B\"ohringer, H. 1996, \mnras,
283, 690.
\bibitem[Ponman et al.(1999)Ponman, Cannon, Navarro]{ponman99}
Ponman, T. J., Cannon, D. B., \& Navarro, J. F. 1999, \nat, 363, 51.
\bibitem[Shu(1992)]{shu92}
Shu, F. 1992, "Gas Dynamics" (University Science Books).
%\bibitem[Spitzer(1978)]{spitzer78}
%Spitzer, L. 1978, "Physical Processes in the Interstellar Medium" (New York:
%Wiley).
\bibitem[Thornton et al.(1998)]{thornton98}
Thornton, K., Gaudlitz, M., Janka, H.-TH., \& Steinmetz, M. 1998, \apj, 500,
95.
\bibitem[Tozzi et al.(1999)Tozzi, Scharf, \& Norman]{tozzi99}
Tozzi, P., Scharf, C. \& Norman, C. 2000, \apj, 542, 106.
%\bibitem[Tozzi \& Norman(2001)]{tozzi01}
%Tozzi, P., \& Norman, C. 2001, \apj
\bibitem[Valageas \& Silk(1999)]{valageas99}
Valageas, P., \& Silk, J. 1999,  \aap, 350, 725.
\bibitem[Wardle et al.(1998)]{wardle98}
Wardle, J. F. C. et al. 1998, Nature, 395, 457.
\bibitem[Xue \& Wu(2000)]{xue00} Xue, Y.\ \& Wu, X.\ 2000,
\apj, 538, 65
\bibitem[Yamada et al.(1999)Yamada, Sugiyama, & Silk]{yamada99}
Yamada, M., Sugiyama, N., \& Silk, J. 1999, \apj, 522, 66.
\end{thebibliography}
\end{document}